\documentclass[entropy,article,submit,oneauthor,pdftex,12pt,a4paper]{mdpi}

\lastpage{x}
\doinum{10.3390/------}
\pubvolume{xx}
\pubyear{2012}
\history{Received: xx / Accepted: xx / Published: xx}

\Title{On chirality of the vorticity of the Universe}

\Author{Davor Palle}

\address{%
Zavod za teorijsku fiziku, Institut Rugjer Bo\v skovi\' c \\
Bijeni\v cka cesta 54, HR-10000 Zagreb, Croatia \\
email: palle@irb.hr, phone: +385-1-4561030, 
fax: +385-1-4680223
}

\corres{palle@irb.hr, phone: +385-1-4561030, fax: +385-1-4680223}

\abstract{
The presence of dark energy in the Universe challenges
the Einstein's theory of gravity at cosmic scales.
It motivates
the inclusion of rotational degrees of freedom in the
Einstein-Cartan gravity, representing the minimal and
the most natural extension of the General Relativity.
One can, consequently, expect the violation of the
cosmic isotropy by the rotating Universe.
We study chirality of the vorticity of the Universe
within the Einstein-Cartan cosmology.
The role of the spin of fermion species during the evolution of the
Universe is studied by averaged spin densities and Einstein-Cartan
equations.
It is shown
that spin density of the light Majorana neutrinos
acts as a seed for vorticity at early stages of
the evolution of the Universe. Its chirality
can be evaluated in the vicinity of the spacelike 
infinity.
It turns out that vorticity of the Universe has
right-handed chirality. 
}


\begin{document}

\section{Introduction}

The two major problems in cosmology and particle physics,
namely, dark energy and dark matter,
force us to modify substantially our theories of
basic interactions. In this paper, we show how our
BY theory of ref. \cite{Palle1} is compatible and
closely related to the Einstein-Cartan cosmology.

To comprehend all the phenomenology in particle
physics and cosmology one has to construct mathematically
consistent and complete theories based on few basic
physical principles.
The theory presented in ref. \cite{Palle1} is an attempt
of solving two main obstacles in the Standard Model (SM)
and relativistic quantum field theory: zero-distance
singularity and causality-violating $SU(2)$ global
anomaly.

The resulting theory (called BY in ref. \cite{Palle1})
is UV finite (not only renormalizable) with heavy and
light Majorana neutrinos as cold and hot dark matter
\cite{Palle2,Palle3}.
There is a perfect balance between bosonic (electroweak 
gauge bosons) and fermionic (leptons and quarks) particles,
owing to exact cancellation of anomalous effective
actions and the constraint relation between boson and fermion
mixing angles $\theta_{W}=2(\theta_{12}+\theta_{23}
+\theta_{31})$. The left-handed chirally-asymmetric weak
interactions appear as an inevitable consequence of the assumed
dimensionality and noncontractiblity of the physical
spacetime.

The absence of Higgs particles is crucial for the
cosmological stability of heavy Majorana neutrinos 
$\tau_{N_{i}} \gg \tau_{U}$ \cite{Palle2}.
The lepton-number violation, the conservation of
$B-L$, as well as lepton and baryon CP violation, lead
to leptogenesis and baryogenesis.

To summarize, the BY gauge theory is not only mathematically 
superior to the SM, but also phenomenologically: (1) solar, atmospheric,
neutrino and long-baseline neutrino experiments favor massive
light neutrinos with mixed flavors, (2) contrary to the SM,
the BY theory has heavy Majorana neutrinos as cold dark matter candidates,
(3) the SM cannot generate lepto- and baryogenesis while the lepton-number
violation in the BY theory together with CP violating phases in the
quark and lepton mixing matrices allow cosmological lepto- and
baryogenesis, (4) quantum-loop corrections 
    in the electroweak and strong interactions
      in the SM show some deviations for the forward-backward and
      left-right asymmetry form factors measured by LEP2 and SLC and 
      a difference from the QCD amplitudes at the largest momenta measured by
      Tevatron. The BY theory can account for these differences 
\cite{Palle7a,Palle7b},
 however only new LHC data will select the proper symmetry-breaking
      mechanism. The most recent predictions of the BY theory for
strong \cite{Davor1}(t-quark charge asymmetry)
and electroweak interactions (CP violation and rare decays)\cite{Davor2,Davor3}
are testable in the near future at hadron colliders. 

Formulating the theory of the local structure of
spacetime as local $SU(3)\times 
SU(2)\times U(1)$ gauge theories, we choose the Einstein-Cartan
theory as formulated by Sciama and Kibble to be the 
theory of the global structure of spacetime.
Trautman was the first who realized the possibility 
of the nonsingular Einstein-Cartan (EC) cosmology \cite{Trautman}.
In addition, there is more freedom to avoid noncausal
Goedel cosmological solutions \cite{Ray,Obukhov}.
The Einstein-Cartan gravity is also a quantum theory 
of gravity but not in a sense of introducing the spin 2 local
quantum field with the corresponding Heisenberg 
commutation rules. The quantum  principle figures only through
quantum mechanical spin densities at the first quantized level.

In this paper we attempt to make a connection between the chirally
asymmetric weak interactions and possibly anisotropic Universe
described by the Einstein-Cartan cosmology.

\section{Theoretical scenario}

The EC gravity relates rotational degrees of freedom
of matter and spacetime, i.e. total angular momentum
as a conserved quantity in the Special Theory of Relativity
consisting of the orbital angular momentum and spin 
\cite{Bjorkena,Bjorkenb} of matter vs. torsion of spacetime.
Spin, as an internal angular momentum of particles,
is a quantum mechanical quantity, so it vanishes in the
classical limit of the vanishing Planck constant.

One can introduce
the angular momentum in General Relativity only 
as a nonconserved quantity and it does not obey tensorial
transformations \cite{Weinberg}. On the contrary, the EC theory
of gravity incorporates spin and angular momenta of matter
and torsion of spacetime invariantly with respect to the general
coordinate transformations of the enlarged general theory
of relativity \cite{Hehla,Hehlb,Hehlc}.
Owing to the algebraic relation between 
spin and angular momentum vs. torsion, one can incorporate 
spin and angular momentum into the effective energy-momentum
tensor \cite{Ray,Obukhov,Hehla,Hehlb,Hehlc}:

\begin{eqnarray}
R_{\mu\nu}-\frac{1}{2}g_{\mu\nu}R=\kappa T_{\mu\nu}^{eff},
\end{eqnarray}
\begin{eqnarray*}
T_{\mu\nu}^{eff}=-p_{eff}g_{\mu\nu}+u_{\mu}u_{\nu}
(p_{eff}+\rho_{eff})-2(g^{\alpha\beta}+u^{\alpha}u^{\beta})
\nabla _{\alpha}[u_{(\mu}S_{\nu)\beta}], \\
\kappa =8\pi G_{N}c^{-4},\ \rho_{eff}=\rho-\kappa S^{2}+\Lambda,
\ p_{eff}=p-\kappa S^{2}-\Lambda, \\
S^{2}=\frac{1}{2}S_{\alpha\beta}S^{\alpha\beta},\ 
S^{\mu}_{.\ \alpha\beta}=u^{\mu}S_{\alpha\beta},\
(\alpha\beta)=\frac{1}{2}(\alpha\beta+\beta\alpha),
\end{eqnarray*}
\begin{eqnarray*}
torsion=Q^{\mu}_{.\ \alpha\beta},\ 
Q^{\mu}_{.\ ab}+2h^{\mu}_{[a}Q_{b]}=\kappa S^{\mu}_{.\ ab}, \\
Q_{a}=h^{\mu}_{a}Q_{\mu},\ Q_{\mu}=Q^{\nu}_{.\ \mu\nu},\ 
[\mu\nu]=\frac{1}{2}(\mu\nu-\nu\mu), \\
a,b=local\ Lorentzian\ indices,\ h^{\mu}_{a}=tetrad\ basis, \\
\ \eta_{ab}=diag(+1,-1,-1,-1).
\end{eqnarray*}

We denote torsion by $Q^{\mu}_{.\ \alpha\beta}$ and total
angular momentum by $S^{\mu}_{.\ \alpha\beta}$.

At the weak interaction scale
$R_{min}={\cal O}(10^{-16}cm)$ torsion is dominated 
by fermion spin densities \cite{Palle4}, while at the largest scale
$R=\infty$ torsion could be dominated only by the angular momentum
of the whole Universe (galaxies, groups, clusters, ...) \cite{Ray,Obukhov,Palle4}.
The contribution of the
torsion at present ($R_{0}={\cal O}(10^{28}cm)$) is much
smaller than the mass density if the Hubble constant is small
or could be much larger if the Hubble constant is
large \cite{Palle8}. This is a consequence of the strong constraints
from the age of the Universe, inevitable negative contribution 
to the integrated Sachs-Wolfe effect and the observed large 
peculiar velocities of clusters at large scales.
The second scenario with large Hubble constant and large torsion at
present is more probable from the theoretical and observational points
of view \cite{Palle8}.
The contribution of the torsion terms to the effective energy-momentum
tensor is always negative with respect to the mass density.

The primordial mass density contrast is evolved 
from quantum fluctuations of the spin to the value at the photon
decoupling defined by parameters of metric beyond that
of Robertson-Walker \cite{Palle5a,Palle5b,Palle5c}.

The question posed is whether it is possible to generate a vorticity of the Universe
and to fix its chirality?

The answer is positive, provided the local and global 
theories of spacetime are BY and EC theories.

Within this framework 
the evolution scenario is the following:

(1) Assuming CP violation in lepton sector, a dynamics
of heavy Majorana neutrinos produces imbalance between
leptons and antileptons \cite{Sakharov}
before their decoupling from primordial plasma.
An example of leptogenesis generated with a Higgs mechanism and with heavy
leptons can be find in the literature \cite{Luty}, but heavy 
leptons are then cosmologically unstable.
The masses of heavy Majorana neutrinos in the BY theory are 
in the range from ${\cal O}(TeV)$ to ${\cal O}(10^{3}TeV)$, thus
leptogenesis happens arround $T > {\cal O}(10^{3}TeV)$.

(2) Before light Majorana neutrino decoupling 
($T_{dec}=few\ MeV$) the imbalance in the number
of baryons and antibaryons appears as a consequence
of the surplus of leptons against antileptons and
conserved B-L \cite{tHoofta,tHooftb,Kolb}.

(3) Now follows the crucial observation: Part of the survived
leptons, like electron, are produced together with
neutrinos through charged current:$W^{-} \rightarrow e^{-}\nu_{M}$
and the helicity \cite{Bjorkena,Bjorkenb}
of the Majorana neutrino $\lambda (\nu_{M})$ 
is predominantly positive ($\lambda (\nu_{M})=+1$).
This is the consequence of the two facts: (a) helicity of Dirac
antineutrinos (=helicity of Majorana neutrinos) produced in $W^{-} \rightarrow e^{-}
\stackrel{-}{\nu_{D}}$
is positive for left-handed weak interactions and
(b) production of the negative helicity relativistic
Majorana neutrinos is suppressed in the same process with weak charged currents
by the kinematical
factor $\frac{m_{\nu}}{E} \ll 1$ \cite{Palle1,Kayser}.
Note that the ratio of the partial decay widths of weak bosons is
$\Gamma (W^{-} \rightarrow e^{-}\nu)/\Gamma (Z \rightarrow \nu\nu) \simeq 1.35$.
Neutral currents do not generate imbalance in neutrino's helicities, because
one relativistic neutrino of a produced pair has negative and the other one
has positive helicity. Thus, irrespective of the details of the physical processes during
leptogenesis, produced number $n_{+}$ of neutrinos with positive helicity
$\lambda (\nu_{M})=+1$ vs. produced number $n_{-}$ of neutrinos with negative helicity
$\lambda (\nu_{M})=-1$ is roughly $n_{+}/n_{-} = {\cal O}(10)$.

(4) It is easy to estimate number of neutrinos and 
other particles at the epoch of neutrino decoupling
\cite{Palle1,Kolb}:

\begin{eqnarray*}
n_{\nu}(T_{0}=2.73 K)={\cal O}(10^{2}cm^{-3}), T_{dec}(\nu)=few\ MeV,
\\
n_{e^{-}}(T_{0})=n_{p}(T_{0}) \simeq n_{B}
(T_{0})={\cal O}(10^{-7}cm^{-3}), 
\end{eqnarray*}
\begin{eqnarray}
m_{N_{i}}={\cal O}(10 TeV)-{\cal O}(100 TeV),\ 
m_{\nu_{i}}={\cal O}(10^{-3} eV)-{\cal O}(1 eV) \nonumber \\
\Rightarrow 
\frac{n_{\nu}}{n_{e^{-}}},\ 
\frac{n_{\nu}}{n_{N}},\ 
\frac{n_{\nu}}{n_{B}} \gg 1,\ at\ T=few\ MeV.
\end{eqnarray}

We conclude that the spin density of the matter is dominated
by the spin of the light neutrinos with an excess of the positive
helicity states. 
In addition, the sum of the orbital angular
momenta of particles vanishes because of the isotropy of the Universe at
that moment of the evolution.

The averaging procedure for spin of cosmic fluids \cite{Kerlick},
lepton and CP violations, together with the positive
helicity abundance, guarantee the nonvanishing spin, as well as 
nonvanishing torsion because of their
algebraic relations within Einstein-Cartan equations.
This asymmetry is a seed for the small
vorticity with the well defined chirality that we define later.
It is clear that any cosmic observer measures different axis of vorticity,
but a chirality of the vorticity is well defined invariant quantity for any observer.

(5) The particles that can compete with neutrinos in
abundances are the background photons.
However, photons, as a massless gauge boson particles,
do not generate the torsion in a gauge invariant
way \cite{Hayashi}.

(6) Electromagnetic and strong forces, as well as the Riemann
curvature of spacetime induced by the metric alone,
are chirally-symmetric interactions and
cannot alter the chirality
of vorticity in later stages of evolution after 
neutrino decoupling. 

Vorticity induced by spin density within EC cosmology (see equations
below) acts as a seed for vortical motions of cosmic particles and
as a seed for the angular momenta of galaxies, clusters,..., Universe
\cite{Kolb,Peebles,Binney}.
Torsion grows together with a grow of large scale structures
at late stages of the structure formation because the torsion is then
dominated by the orbital angular momenta of large scale structures.

(7) A term of EC equations linear in torsion (angular momentum) allows
us to uniquely determine chirality of the vorticity.
Namely, this is possible to achieve because of the relation
derived at spacelike infinity and its vicinity \cite{Palle4,Penrose}.
The present Universe is in the matter dominated epoch with
$T_{\gamma, 0}=2.73\ K,\ \frac{\rho_{\gamma, 0}}{\rho_{m, 0}}=
{\cal O}(10^{-4})$, while the spacelike infinity is at 
$T_{\gamma}(R=\infty)=0\ K$. Evidently, the present state of the
Universe is in the vicinity of the spacelike infinity.

We define the metric with vorticity 
with two real parameters $m,\Sigma$ \cite{Obukhov,Palle4},
spin(ang. momentum) and torsion of the fluid \cite{Obukhov}, as well as the
standard definition of vorticity \cite{Palle4}.
The crucial relations (eqs.(3)) at spacelike infinity allows to
fix the chirality of the vorticity from a chirality of the angular
momentum:

\begin{eqnarray*}
by\ definition:\ 
S^{\alpha\beta}=-\frac{1}{2}n h^{\alpha}_{i}h^{\beta}_{j}\mu^{ij},
\end{eqnarray*}
\begin{eqnarray*}
metric:\ ds^{2}=dt^{2}-dx^{2}-(1-\Sigma)e^{2mx}dy^{2}-dz^{2}
-2\sqrt{\Sigma}e^{mx}dydt, \\
\Sigma,\ m=constant\ parameters, 
\end{eqnarray*}
\begin{eqnarray}
R=\infty :\ 
Q=Q^{\hat{0}}_{.\ \hat{1}\hat{2}}=-\frac{m(2-\Sigma)}{2\sqrt{\Sigma}},   \   \Sigma={\cal O}(10^{-3})\ll 1\ ,  
\end{eqnarray}
\begin{eqnarray*}
G_{N}\rho_{\infty}H^{-2}_{\infty}=\frac{3}{4 \pi},\ H_{\infty}=Hubble\ constant\ at\ \infty, \\
by \ definition:\ Q=\frac{1}{2}\kappa n\mu^{\hat{2}\hat{1}}B_{\infty},
\ \mu^{\hat{1}\hat{2}}=-\mu^{\hat{2}\hat{1}}=+\frac{1}{2}\hbar, \\
n=n_{\nu}(\lambda=+1)-n_{\nu}(\lambda=-1) > 0 .
\end{eqnarray*}

The asymmetry in the lepton or baryon numbers is large 
from the beginning of the activation of CP violating processes by weak 
interactions.
$B_{\infty}$ is some positive amplification factor as a result of the evolution.
Its magnitude can be estimated $B_{\infty}\simeq Q(Universe)_{0}/
\kappa Spin(neutrinos)_{0}={\cal O}(10^{34})$. This is a consequence
of the fact that the torsion of the Universe $Q(Universe)_{0}$ at
present time is predominantly defined by its macroscopical orbital
angular momentum and it cannot vanish because of the nonvanishing
primordial vorticity. From preceding equations and ref. \cite{Palle4},
it follows $|Q_{0}| \simeq \sqrt{3} H_{0}$, therefore the orbital
angular momentum of the Universe can be estimated to be
$L_{U} \simeq \frac{1}{\sqrt{3}} G_{N}^{-1}H_{0}^{-2}=
2 h^{-2}\times 10^{94} g cm^{2} s^{-1}$.

It is advantageous to check the estimate of the angular momentum
of the Universe by the relation of the rotational support
\cite{Paddy}
(rotation against gravity) $L = \sigma \frac{G_{N}M^{5/2}}{|E|^{1/2}}$,
where $\sigma$ is a dimensionless spin parameter, M the mass and
E the energy of the physical system $|E| \simeq \frac{G_{N}M^{2}}{R}$.
Acknowledging that $R_{U} = H_{0}^{-1}$, $M_{U}=\Omega_{m} \rho_{c} 
\frac{4 \pi}{3} R_{U}^{3} = \frac{\Omega_{m}}{2} G_{N}^{-1} H_{0}^{-1}$,
one easily gets $L_{U} = \sigma (\frac{\Omega_{m}}{2})^{3/2}
G_{N}^{-1} H_{0}^{-2}$. Rotational support requires that
$\sigma = {\cal O}(1)$, while in the EC cosmology $\Omega_{m}=2$  
\cite{Palle4}, thus the estimate for $L_{U}$ agrees with a 
previous one derived in the EC cosmology.

Now, we acknowledge the fact of 
the abundant positive-helicity states of neutrinos contributing to the
spin density and leading to the positive chirality angular momentum 
of the Universe.
It follows then from eqs.(3) 

\begin{eqnarray*}
Q < 0 \Rightarrow m > 0 .
\end{eqnarray*}

Definition and chirality of vorticity are naturally defined
as \cite{Som,Ehlers,Kreyszig} (standard right-handed xyz frame is
assumed):

\begin{eqnarray*}
\omega_{\nu\mu}&=&\frac{1}{2}(\nabla_{\alpha}u_{\beta}
-\nabla_{\beta}u_{\alpha})P^{\alpha}_{\mu}P^{\beta}_{\nu},\\
P_{\alpha\beta}&=& g_{\alpha\beta}-u_{\alpha}u_{\beta},\ 
\omega_{ij}=h^{\mu}_{i}h^{\nu}_{j}\omega_{\mu\nu}, \\
\omega^{\nu\mu}&=&h^{\nu}_{i}h^{\mu}_{j}\omega^{ij},
\end{eqnarray*}
\begin{eqnarray*}
\Sigma H_{\infty} \simeq \frac{2}{\sqrt{3}}\omega_{\infty},\ 
\omega^{2}_{\infty}=\frac{1}{2}\omega_{\mu\nu}\omega^{\mu\nu},
\end{eqnarray*}
\begin{eqnarray}
\omega_{\mu\nu}(m=+|m|,xyz)=-\omega_{\mu\nu}(m=-|m|,xyz)
=-\omega_{\mu\nu}(m=+|m|,yxz), \nonumber \\
m > 0\ and\ \omega_{\hat{1}\hat{2}}=\omega^{\hat{1}\hat{2}}
=+m\frac{\sqrt{\Sigma}}{2} > 0 \Rightarrow right-handed\
vorticity .
\end{eqnarray}

Hence, if the abundant positive-definite helicity light neutrinos
define the torsion's chirality, then the chirality of the vorticity
of the Universe is right-handed, i.e. it is positive in the standard
right-handed frame.

\section{Conclusions and remarks}

The magnitude and chirality of the vorticity and angular
momentum of the Universe, as well as the relic-neutrino helicities,
are well defined observables. Numerous statistical studies of the
WMAP data reported a violation of the isotropy of the Universe,
while analyses of clusters found anisotropic large scale
flows (see refs. in \cite{Palle8}).

Dark energy is, in our scenario, the angular momentum of the Universe,
giving the negative contribution to the effective mass
density of the Universe (from eqs. (3): $\Omega_{m}=2,
\Omega_{Q}=-1$ and $\Omega_{\Lambda}=0$, because
$\lim_{R\rightarrow \infty} \rho_{\Lambda} =
-\frac{1}{2}\lim_{R\rightarrow \infty} \rho_{m} = 0$).
Note that in the quasilinear regime, at the galaxy forming epoch,
$\rho_{Q} \propto (1+z)^{3}$ \cite{Palle8}, while in the
nonlinear galaxy-cluster forming epoch (at very low redshifts),
one expects $\rho_{Q} \propto const$ \cite{Paddy}.

The existence of the lepton CP violation is indispensable
\cite{Sakharov}
in our cosmological scenario with BY and EC theories.
Recent results of MiniBooNE and MINOS experiments with different
oscillation results for neutrinos and antineutrinos strongly suggest
possible lepton CP violation if the data are fitted with three neutrino
flavors.
Note that the CP violating phase need not to be accompanied
with $\theta_{31}$ mixing angle, but can be attached
to $\theta_{12}$ mixing angle. The inclusion of large CP violating
phase can substantially change the present estimate of the mixing
angles.

We see that the chirality of the asymmetry in particle physics
is left-handed (weak interactions), thus the opposite to the positive
chirality of the vorticity of the Universe, all measured in the
right-handed coordinate reference frames. If we change a reference
frame to be left-handed, the chiralities change signs. The sum of
chiralities of the
microstructure and the macrostructure of the physical world
remains zero irrespective of our choice of the reference frame.
Parity violation in particle physics and cosmology is mandatory
from both physical and mathematical points of view.
The PLANCK mission can give us a definitive answer on the 
chirality of the vorticity of the Universe.

\end{document}